# Study of pollution in the El Jadida-Safi Atlantic coastal zone (Morocco) by using PIXE and SSNTD methods

H. Erramli[1*], H.D. Gogon[1], M.A. Misdaq[1], T. Sauvage[2], C. Ramboz[3]

*1. Nuclear Physics and Techniques Laboratory, Faculty of Sciences Semlalia, University Cadi Ayyad, BP. 2390, Marrakech, Morocco*

*2. CNRS/CEMHTI, site cyclotron, 3A, rue de la Férollerie, F 45071 Orléans Cedex 2, France*

*3. Institut des Sciences de la Terre d'Orléans (ISTO, CNRS - 1A, rue de la Férollerie, F 45071 Orléans Cedex 2, France*

## Abstract

In this work PIXE experiments were performed for measuring heavy and light elements (ranging from aluminium to lead) concentrations inside various polluted and unpolluted soils as well as liquid samples collected from different phosphate factory sewers in the El Jadida-Safi Atlantic coastal region (Morocco). In addition, uranium ($^{238}$U) and thorium ($^{232}$Th) contents were evaluated in the same samples studied by using CR-39 and LR-115 type II solid state nuclear track detectors (SSNTDs). The influence of the phosphate industry wastes on the concentrations of both radioactive and non-radioactive elements of the samples studied was investigated.

* Corresponding author. Tel.: +212 24 43 46 49;fax: +212 24 43 74 10
E-mail address: hassane@ucam.ac.ma or erramli@yahoo.com (H. Erramli)

# 1. Introduction

The Moroccan Office Chérifien des Phosphates (OCP) is a national company in load of all phosphate products. It is the second phosphate productive enterprise in the world, very little after American IMC Agrico Co. (Group Freeport McMoran). Mining is carried out on three main sites: Khouribga, Gantour and Boucraâ. The centres of transformation of phosphate into phosphoric acid are established in the littoral industrial park of Jorf Lasfar and Safi. The OCP is by far the first world exporter of rough phosphates and phosphoric acid and one of the largest international suppliers of fertilizers such as triple superphosphate (TSP) and diammonium phosphate (DAP).

Internationally, interest of pollution in urban and global environments has increased significantly over the last decade (Cohen, 1998 ; USEPA Report 1999). Physicists, geologists, archaeologists, art conservators and others have utilised many analytical techniques for the determination of the compositions of polluted samples in environment, soil, rocks and minerals, and they continue to investigate emerging technologies for their utility. Their interest arises from the fact that a geologist can deduce information about the physical and chemical conditions under which a material was formed and through which the material has existed, from knowledge of soil, rocks and minerals composition. Specific materials are defined by their major elements, but it is the amount and distribution of the minor or trace elements that are more indicative of the materials geochemical environment. Thus analytical techniques, with more indicative limits, are of interest. Proton Induced X-ray Emission (PIXE) analysis used for more than thirty years, is a powerful yet non-destructive elemental analysis technique and is a promising tool for the study of trace element behaviour in a wide variety of materials (geological, archaeological, biological ….). The combination of a reasonable resolution of a few microns combined with a detection power in the ppm ($10^{-6}$g.g$^{-1}$) range offers possibilities for trace element mapping giving insight in trace element composition. PIXE method has been widely used for trace elemental analysis since Johansson et al. (1970) achieved mass detection limits in the region of $10^{-12}$g. The extensive literature on the subject includes several excellent review articles (Johansson S.A.E. and Johansson T.B.,1976 ; Owers and Shalgosky, 1974).

In the present work, the PIXE and SSNTD methods were employed to measure heavy and light elements in various material samples collected from different sites of the El Jadida-Safi Atlantic coastal region (Morocco).

## 2. Experimental methods

Six solid samples (S1, S2, S3, S4, S5 and S6) were collected from different sites in the El Jadida-Safi Atlantic coastal region (Morocco) (Fig.1), pounded and homogenized. S1 was collected from a deposit of raw sulphur material in the Jorf Lasfar port, S2 and S6 were directly collected from two dumps of recent phosphate wastes resulting from the phosphate industrial activities in the study area, S3 and S4 were collected from the El Oualidia and Sidi Moussa agricultural zones and S5 was collected from the beach of the Safi city. Eight water samples were directly collected from the sewers of the phosphate factories situated in the same study area (Fig.1) and filtered. The resulting residues "foams": WR1, WR2, WR3, WR4, WR5, WR6, WR7 and WR8 were dried.

Almost 2 g of each soil and water residue sample were compacted in a pellet of 1.5 cm diameter and few mm thickness using a mechanical press. The obtained pellets were stuck on an aluminium disc. The prepared soil and water residue samples were then analysed by PIXE method. The PIXE experiments were performed at the CERI-CNRS, Orléans, using a 3.5 MV HVEC Van De Graaff accelerator. The miniprobe focuses a 2.5 MeV proton beam to a target placed under vacuum (from $5x10^{-5}$ to $10^{-6}$ mbar). Detail descriptions have been reported previously by Zine et al. (1990) and by Choï (1996). The X-ray Si(Li) detector (Oxford instruments) is installed at a 135° angle with respect to the beam axis and is 2.4 cm distant from the sample. This detector is characterized by a 30 $mm^2$ nominal surface area, a 3 mm nominal Si-crystal thickness and 7.5 μm-thick Be window. Its energy resolution at 5.9 keV is 148 eV. The dead time of the Si(Li) detector is of 25 μs. A 135 μm-thick Be-filter is placed in front of the detector to prevent interactions with scattered protons. A 200 μm thick Al-funny filters and bored of a hole of 0.77 mm is superimposed on the Be-filter in order to attenuate the characteristic X-rays from major elements, which would disturb the electronic detection, increase pulse pile-up on the spectra and obliterate X-rays of trace elements. A mirror, placed in front of the target, reflects the image of the beam impact on the target to endoscopes, which magnifies at 200x (Gama et al., 2001). Standards and samples are polished and coated with a thin carbon layer in order to ensure conductivity. We used a beam current of 0.8 nA keeping the counting rate lower than $1x10^3$ counts.$s^{-1}$ to measure major and minor elements. The intensity of the proton beam is measured using a chopper calibrated relative to a Faraday cup. The relative uncertainty of charge integration is around 5 %. The typical beam spot size was about 0.5x0.5 $mm^2$. The execution time for our samples was around 20-30 min, depending on the count rate, but the actual time needed for a routine test may be shorter. For each sample, characteristic X-rays produced from the proton bombardment at four different positions on the sample surface were measured. A typical PIXE spectrum and its corresponding GUPIX fit (Campbell and Maxwell, 1996) are shown in Fig.2.

Disk shaped CR-39 (manufactured by Pershore Mouldings Ltd., U.K) and LR-115 type II (manufactured by Kodak Pathé, France and marketed by Dosirad, France) solid state nuclear track detectors (SSNTDs) of radius q =2 cm have been separately placed in close contact with a soil/water residue sample in a hermetically sealed (using a glue) cylindrical plastic container for one month (30 days) [12]. During this exposure α-particles emitted by the nuclei of the uranium-238 and thorium-232 series bombarded the SSNTD films. After the irradiation, the exposed films were etched in a NaOH solution at optimal conditions of etching, ensuring good sensitivities of the SSNTDs and a good reproducibility of the registered track density rates: 2.5 N at 60°C for 120 minutes for the LR-115 type II films and 6.25 N at 70 °C for 7 hours for the CR-39 detectors (Misdaq et al., 2000). After this chemical treatment, the track densities registered on the CR-39 and LR-115 type II SSNTDs were determined using an optical microscope. Backgrounds on the CR-39 and LR-115 type II SSNTDs were evaluated by placing these films in empty well-closed plastic containers identical to those used for analysing honey samples for one month and counting the resulting track densities. This operation was repeated ten times: track densities registered on the CR-39 and LR-115 type II detectors were found to be identical within the statistical uncertainties. As the system is well-sealed (there is no escape of radon and thoron) and the exposure time was 30 days, one can assume radioactive secular equilibrium between uranium, thorium and their corresponding decay products. For our experimental etching conditions, the residual thickness of the LR-115 type II detectors measured by means of a mechanical comparator is 5 μm. This thickness defines the lower ($E_{min}$ = 1.6 MeV) and upper ($E_{max}$ = 4.7 MeV) energy limits for registration of tracks of α-particles in LR-115 type II films (Misdaq et al., 2000). All α-particles emitted by the uranium and thorium series that reach the LR-115 detector at an angle lower than its critical angle of etching $\theta'_c$ with a residual energy between 1.6 MeV and 4.7 MeV are registered as bright track-holes. The CR-39 detector is sensitive to all α-particles reaching its surface at an angle smaller than its critical angle of etching $\theta_c$. $\theta'_c$ and $\theta_c$ were calculated using a method described in detail by Misdaq et al.(1999).

The global track density rates (tracks.cm$^{-2}$.s$^{-1}$), due to α-particles emitted by the uranium and thorium series inside a material sample, registered on the CR-39 and LR-115 type II detectors, after subtracting the corresponding backgrounds, are respectively given by (Misdaq et al., 2000) :

$$\rho_G^{CR} = \frac{\pi q^2}{2 S_d} C(U) d_s \left[ A_U \sum_{j=1}^{8} k_j R_j \varepsilon_j^{CR} + A_{Th} \frac{C(Th)}{C(U)} \sum_{j=1}^{7} k_j' R_j' \varepsilon_j^{'CR} \right] \quad (1)$$

and

$$\rho_G^{LR} = \frac{\pi q^2}{2 S_d'} C(U) d_s \left[ A_U \sum_{j=1}^{8} k_j R_j \varepsilon_j^{LR} + A_{Th} \frac{C(Th)}{C(U)} \sum_{j=1}^{7} k_j' R_j' \varepsilon_j^{'LR} \right] \quad (2)$$

where $S_d$ and $S_d'$ are respectively the surface areas of the CR-39 and LR-115 type II films, $C(U)$ (µg.g$^{-1}$) and $C(Th)$ (µg.g$^{-1}$) are the uranium ($^{238}$U) and thorium ($^{232}$Th) concentrations of the sample, $A_U$(Bq.g$^{-1}$) = 0.0123 and $A_{Th}$(Bq.g$^{-1}$) = 0.0041 are the specific activities of a sample for a $^{238}$U content of 1ppm and a $^{232}$Th content of 1ppm, $d_S$ is the density of the sample (g.cm$^{-3}$), $R_j$ and $R_j'$ are the ranges, in the sample, of an α-particle of index j and initial energy $E_j$ emitted by the nuclei of the uranium and thorium series, respectively, $k_j$ and $k_j'$ are respectively the branching ratios corresponding to disintegration of the nuclei of the uranium and thorium series and $\varepsilon_j^{CR}$, $\varepsilon_j^{'CR}$, $\varepsilon_j^{LR}$ and $\varepsilon_j^{'LR}$ are respectively the detection efficiencies of the CR-39 and LR-115 type II detectors for the emitted α-particles. The first terms (right of Eqs. (1) and (2)) correspond to the number of α-particles emitted by the uranium family (8α-emitting nuclei), whereas the second terms correspond to the number of α-particles emitted by the thorium series (7α–emitting nuclei).

Combining Eqs. (1) and (2), we obtain the following relationship between track density rates and thorium to uranium ratios :

$$\frac{C(Th)}{C(U)} = \frac{A_U}{A_{Th}} \times \frac{\frac{S'_d}{S_d}\sum_{j=1}^{8}k_j\varepsilon_j^{CR}R_j - \frac{\rho_G^{CR}}{\rho_G^{LR}}\sum_{j=1}^{8}k_j\varepsilon_j^{LR}R_j}{\frac{\rho_G^{CR}}{\rho_G^{LR}}\sum_{j=1}^{7}k'_j\varepsilon_j^{'CR}R'_j - \frac{S'_d}{S_d}\sum_{j=1}^{7}k'_j\varepsilon_j^{'LR}R'_j} \quad (3)$$

The uranium content of a soil/water residue sample is given by (Eq.(2)) :

$$C(U) = \frac{2S'_d \rho_G^{LR}}{\pi q^2 d_s \left[ A_U \sum_{j=1}^{8}k_j\varepsilon_j^{LR}R_j + \frac{C(Th)}{C(U)} A_{Th} \sum_{j=1}^{7}k'_j\varepsilon_j^{'LR}R'_j \right]} \quad (4)$$

By calculating first the detection efficiencies of the CR-39 ($\varepsilon_j^{CR}$ and $\varepsilon_j^{'CR}$) and LR-115 type II ($\varepsilon_j^{LR}$ and $\varepsilon_j^{'LR}$) SSNTDs for α-particles emitted by the thorium-232 and uranium-238 series inside a soil/water residue sample (Misdaq et al., 2000), and secondly by measuring track density rates (tracks.cm$^{-2}$.s$^{-1}$) registered on the CR-39 ($\rho_G^{CR}$) and LR-115 type II ($\rho_G^{LR}$) films one can evaluate the $^{238}$U and $^{232}$Th contents inside the considered material sample.

## 3. Results and discussion

For our PIXE analysis, the beam spot was directed at the surface of sample, and four measurements were made for each specimen. The extensive range of elements from Al to Pb is clearly visible, including many of the key metals of interest. The spectrum consists of a number of peaks corresponding to the $K_\alpha$ and $K_\beta$ X-rays due to several chemical elements in a sample (Fig.2). For the heaviest elements the cross section for K X-ray production is very small but instead the L X-rays turn up in the spectrum. The peaks are superimposed upon a continuous background originating mainly in the backing material. The first step in determining the element concentrations was to obtain the matrix composition from the Electron Probe Micro-Analyses (EPMA) (work done at BRGM-CNRS-University of Orléans, France). With this matrix composition and from the measured PIXE spectrum, the chemical composition (Z ≥ 13) in the sample could be obtained using a de-convolution program GUPIX (Campbell and Maxwell, 1996).

Different standards were utilized for calibrating our PIXE experiments (a list of the chemistry, name and origin of all standards is given in Table 1). Table 2 shows data obtained for trace and major elements for the fluoroapatite $Ca_5(PO_4)_3F$ (a) and the BR standard (c). Good agreement was found between data obtained by using the PIXE method and the certified values. The statistical limits of detection (LOD) values for elements analysed by means of their K x-ray emission spectra decreased from 30 µg/g for $20 < Z < 35$ to 10 µg/g for $35 < Z < 55$ and then increased again to 50 µg/g for $Z>55$. For the $60<Z<92$ elements analysed by means of their L x-ray emission series, the LOD value is about 50 µg/g.

Data obtained for the major (Si, P, S, Cl, K, Ca, Ti and Fe) and trace (Cr, Ni, Cu, Zn, As, Br, Sr, Y, Zr, Ag, Hg and Pb) elements concentrations of the studied samples are given in Tables 3 and 4, respectively. Figs.3 and 4 show the distribution of major and trace elements in the material samples studied. We noted that the S1, WR2, WR3, WR4, WR5, WR6 and WR8 samples present higher sulphur concentrations than the other samples. This is because:

- The S1 sample was collected from a deposit of raw sulphur material in the Jorf Lasfar port.
- The WR2, WR3, WR4, WR5 and WR6 water residue samples were collected from the sewers of the Jorf Lasfar phosphate factory. Indeed, sulphur is intensively used in the transformation of phosphates.
- The WR8 sample was collected from the sewer of the Safi phosphate factory.

The S2, S6 and WR7 samples show larger phosphorus concentrations than the other samples. This is due to the fact that these samples were collected from phosphate waste dumps and a sewer of the Jorf Lasfar phosphate factory, respectively. The S2, S4, S5, S6 and WR7 samples contain more calcium than the others. This is because:

- The S2 and S6 samples belong to phosphate waste dumps.
- The S4 sample was collected from the El Oualidia agricultural area in which farmers used phosphate fertilizers to increase their agriculture production.
- The S5 sample was collected from the beach of the city of Safi which is polluted by phosphate dusts because it is situated near a phosphate factory.
- The WR7 sample was collected from the sewer of the Jorf Lasfar phosphate factory.

We noted that the WR8 sample corresponding to a water sample collected from the sewer of the Safi phosphate factory contains higher Cu, Zn, As and Hg concentrations than the other water residue samples collected from the sewers of the Jorf Lasfar phosphate factory. The concentration of lead (Pb) is clearly higher in the WR2 water residue sample collected from the sewers of the Jorf Lasfar phosphate factory than in the other samples.

U and Th concentrations were evaluated in the studied soil and water residue samples (Table 5). The relative uncertainty on the U and Th concentration determination was about 8 %. It is to be noticed that our SSNTDs' method was already validated by many instrumental techniques such as Isotope Dilution Mass Spectrometry (IDMS), Neutron Activation Analysis (NAA) and gamma-ray spectrometry for liquid and solid material samples (Misdaq et al., 2000). We noted from results shown in Table 5 that:

- The S2, S4, S5 and S6 soil samples contain higher U concentrations than the S1 and S3 ones. This is because the former samples were polluted by raw phosphate confirmed by their higher Ca and/or P percentage (Table 3) due to the presence of calcite ($CaCO_3$) and apatite ($Ca_5(PO_4)_3(OH,F)$) in these samples.
- There exist two uranium contamination sources for the studied water residue samples; a contamination due to only phosphate wastes (case of the WR7 which presents higher Ca and P contents) (Table 3) and a pollution due to the phosphoric acid processing wastes (case of WR2, WR3, WR5 and WR8 which present higher S content) (Table 3).

## 4. Conclusion

It has been shown by this study that by combining the PIXE method with a Solid State Nuclear Track Detectors' technique one can evaluate the contents of light as well as heavy elements inside various solid and liquid material samples. It has been shown that pollution due to Cu, Zn, As and Hg is more important in the Safi phosphate industrial site than in the Jorf Lasfar one whereas pollution due to Pb is more important in the latter site than in the former one. A good correlation was found between S, Ca, P and U contents of the studied samples. It has been shown that the uranium concentration increase was due to both raw phosphate and phosphate industry wastes in the study area.

**Acknowledgement:**
This work was realized under a CNRST (Morocco)-CNRS (France) research convention

**Table and figure captions**

**Table 1**. Nature and origin of the standard materials used for the PIXE experiments calibration

**Table 2**. Comparison between certified values and data obtained by PIXE for trace and major elements for the fluoroapatite ($Ca_5(PO_4)_3F$) (a) and natural basalt (BR)(b) standards

**Table 3**. Data obtained by using PIXE for the concentrations (in % by weight) of major elements in solid and liquid samples collected from the Atlantic coastal region (Morocco), with zero referring to concentrations below the limits of detection

**Table 4.** Values of the concentrations (in µg/g) of trace elements in the studied material samples obtained by PIXE, with zero referring to concentrations below the limits of detection.

**Table 5.** Data obtained for uranium and thorium concentrations (in µg/g) for the studied material samples obtained by using the SSNTD method

**Fig. 1**. The geographical situation of the study area

**Fig. 2**. Typical PIXE spectrum obtained for the S1 sample

**Fig. 3**. Distribution of major (a) and trace (b) elements in the solid samples studied

**Fig. 4**. Distribution of major (a) and trace (b) elements in the water residue samples studied

| Name, origin | Standard |
|---|---|
| C320, Oxford Instruments | $Li_2Ta_2O_6$, $B_2O_3$, $NaAlSi_2O_6$, $Mg_2SiO_4$, $Al_2SiO_5$, $SiO_2$, $KAlSi_3O_8$, $Ca_5(PO_4)_3F$, $TiO$, $FeS$, $FeCr_2O_4$, $NiO$, $SrTiO_3$, $Nb_2O_5$, $CdSe$, $BaAl_2Si_2O_8$, $Gd_3Ga_5O_{12}$, $PbTe$ |
| BR (88GOV1) SARM Laboratory (Nancy, France) | Natural basalt |

**Table 1**

| Element | Certified values (in % by weight) | PIXE method |
|---|---|---|
| Si | 0.42±0.01 | 0.38±0.04 |
| P | 17.9±0.4 | 17.8±0.2 |
| Cl | 0.020±0.001 | 0.010±0.001 |
| Ca | 38±1 | 38±1 |
| La | 0.27±0.01 | 0.20±0.01 |
| Ce | 0.72±0.02 | 0.73±0.02 |
| Pb | 0.037±0.002 | 0.032±0.001 |

**Table 2(a)**

| Major element | Certified values (in % by weight) | PIXE method |
|---|---|---|
| P | 0.46±0.02 | 0.44±0.02 |
| K | 1.16±0.04 | 1.33± 0.02 |
| Ca | 9.86 ±0.30 | 9.88 ±0.01 |
| Ti | 1.56±0.05 | 1.56 ±0.05 |
| Fe | 9.0 ±0.3 | 9.2± 0.1 |

| Trace element | Certified values (in µg/g) | PIXE method |
|---|---|---|
| P | 4600±140 | 4361±200 |
| K | 11600±350 | 13266±50 |
| Ca | 98600 ±690 | 98832±120 |
| Ti | 15600±51 | 15546±110 |
| Fe | 90000 ±630 | 91798±120 |
| V | 235±7 | 255±25 |
| Cr | 380±11 | 380±16 |
| Mn | 1550±50 | 1384±23 |
| Ni | 260±8 | 266±12 |
| Ga | 19±1 | 18±2 |
| Sr | 1320±40 | 1600±15 |
| Y | 30±1 | 27±3 |
| Zr | 260±8 | 275±14 |
| Nb | 98±3 | 117±6 |
| Ba | 1050±32 | 908±120 |

**Table 2(b)**

| Sample | Si | P | S | Cl | K | Ca | Ti | Fe |
|---|---|---|---|---|---|---|---|---|
| S1 | 20.5±0.2 | 0 | 45.7±0.1 | 0.40±0.02 | 0.36±0.01 | 4.00±0.01 | 0.340±0.005 | 1.80±0.02 |
| S2 | 10.7±0.3 | 21.0±0.2 | 2.81±0.02 | 0.63±0.01 | 0.030±0.003 | 39.61±0.01 | 0.13±0.01 | 0.30±0.01 |
| S3 | 23.6±0.1 | 1.2±0.1 | 0.2±0.02 | 18.50±0.04 | 1.24±0.02 | 26.40±0.04 | 0.10±0.01 | 0.70±0.01 |
| S4 | 4.5±0.2 | 0.90±0.06 | 0.50±0.01 | 1.30±0.01 | 5.1±0.2 | 41.20±0.03 | 0.20±0.01 | 0.32±0.01 |
| S5 | 2.4±0.2 | 0.4±0.1 | 0.32±0.01 | 0.90±0.01 | 5.3±0.2 | 38.1±0.1 | 0.13±0.01 | 0.51±0.01 |
| S6 | 9.6±0.1 | 10.4±0.1 | 1.00±0.01 | 1.44±0.01 | 0.40±0.01 | 29.00±0.02 | 0.12±0.01 | 2.52±0.01 |
| WR1 | 27.0±0.4 | 2.7±0.2 | 6.8±0.1 | 0.27±0.02 | 0 | 9.20±0.04 | 0.020±0.002 | 0.10±0.01 |
| WR2 | 15.0±0.1 | 3.2±0.1 | 16.55±0.03 | 3.00±0.02 | 0.45±0.01 | 25.52±0.02 | 0.20±0.01 | 1.15±0.01 |
| WR3 | 2.4±0.1 | 0.20±0.02 | 38.50±0.05 | 8.20±0.03 | 0.42±0.01 | 17.64±0.02 | 0.36±0.02 | 0.44±0.01 |
| WR4 | 6.3±0.3 | 0 | 37.8±0.2 | 3.15±0.04 | 0 | 24.23±0.05 | 0.020±0.002 | 0.060±0.005 |
| WR5 | 0.60±0.06 | 0 | 41.00±0.1 | 0.42±0.02 | 0 | 25.70±0.03 | 0.07±0.01 | 0.020±0.002 |
| WR6 | 2.0±0.1 | 0 | 39.3±0.1 | 0.50±0.02 | 2.0±0.1 | 24.52±0.03 | 0.060±0.005 | 0.020±0.002 |
| WR7 | 9.4±0.2 | 24.2±0.2 | 1.65±0.02 | 1.04±0.01 | 0.100±0.005 | 48.45±0.04 | 0.060±0.004 | 0.52±0.01 |
| WR8 | 6.2±0.1 | 0.9±0.1 | 7.50±0.02 | 0.64±0.01 | 0.600±0.005 | 6.44±0.01 | 0.200±0.003 | 51.31±0.04 |

**Table 3**

| Sample | Cr | Ni | Cu | Zn | As | Br | Sr | Y | Zr | Ag | Hg | Pb |
|---|---|---|---|---|---|---|---|---|---|---|---|---|
| S1 | 0 | 0 | 126±9 | 263±8 | 0 | 0 | 371±9 | 19±2 | 90±8 | 0 | 0 | 477±15 |
| S2 | 0 | 787±108 | 1200±22 | 266±17 | 0 | 0 | 1095±80 | 935±22 | 0 | 0 | 0 | 293±41 |
| S3 | 137±11 | 0 | 0 | 0 | 0 | 117±5 | 772±14 | 114±8 | 2749±33 | 0 | 0 | 0 |
| S4 | 0 | 219±15 | 570±23 | 146±12 | 33±3 | 0 | 2329±20 | 21±2 | 30±2 | 13±1 | 0 | 204±25 |
| S5 | 0 | 571±44 | 410±24 | 186±15 | 34±3 | 45±4 | 1768±16 | 0 | 185±19 | 0 | 54±4 | 101±10 |
| S6 | 380±26 | 260±30 | 232±8 | 1594±30 | 124±17 | 84±8 | 2016±20 | 126±11 | 30±2 | 144±13 | 88±9 | 552±25 |
| WR1 | 0 | 0 | 0 | 0 | 0 | 0 | 268±21 | 407±28 | 0 | 0 | 0 | 0 |
| WR2 | 425±50 | 0 | 4460±75 | 547±50 | 265±40 | 281±30 | 1059±35 | 650±37 | 217±22 | 185±19 | 0 | 1700±124 |
| WR3 | 3617±186 | 500±46 | 1332±23 | 255±11 | 158±6 | 653±10 | 810±12 | 384±10 | 65±6 | 197±20 | 0 | 58±5 |
| WR4 | 0 | 0 | 83±8 | 51±5 | 0 | 47±10 | 463±23 | 280±20 | 67±6 | 0 | 0 | 0 |
| WR5 | 14140±404 | 0 | 72±7 | 45±4 | 0 | 0 | 495±9 | 91±6 | 31±2 | 0 | 0 | 0 |
| WR6 | 13131±381 | 0 | 2660±22 | 176±9 | 61±6 | 0 | 514±10 | 131±7 | 0 | 26±2 | 0 | 226±18 |
| WR7 | 0 | 640±60 | 365±15 | 207±10 | 0 | 0 | 1049±12 | 407±10 | 0 | 0 | 66±6 | 144±21 |
| WR8 | 0 | 0 | 10856±76 | 1537±51 | 861±27 | 0 | 431±18 | 44±4 | 112±16 | 52±5 | 366±37 | 87±9 |

**Table 4**

| Sample | $\rho_G^{CR} \times 10^{+5}$ (tr.cm$^{-2}$.s$^{-1}$) | $\rho_G^{LR} \times 10^{+5}$ (tr.cm$^{-2}$.s$^{-1}$) | C(U) (μg g$^{-1}$) | C(Th) (μgg$^{-1}$) |
|---|---|---|---|---|
| S1 | 7.5 ± 0.3 | 2.3 ± 0.1 | 4.8 ± 0.2 | 3.1 ± 0.2 |
| S2 | 4.4 ± 0.2 | 1.4 ± 0.1 | 17 ± 1 | 11 ± 0.6 |
| S3 | 2.2 ± 0.1 | 0.70 ± 0.03 | 10.8 ± 0.6 | 15 ± 1 |
| S4 | 6.5 ± 0.3 | 2.0 ± 0.1 | 15 ± 1 | 9.4± 0.5 |
| S5 | 1.7 ± 0.1 | 0.54 ± 0.02 | 19 ± 1 | 14 ± 1 |
| S6 | 2.7 ± 0.1 | 0.85 ± 0.04 | 17 ± 1 | 19 ± 1 |
| WR2 | 0.16 ± 0.01 | 0.050 ± 0.001 | 21 ± 1 | 13.2 ± 0.7 |
| WR3 | 0.30 ± 0.01 | 0.090 ± 0.002 | 25 ± 1 | 19 ± 1 |
| WR5 | 0.22 ± 0.01 | 0.070± 0.001 | 9.2 ± 0.5 | 6.4 ± 0.3 |
| WR7 | 4.0± 0.2 | 1.20 ± 0.05 | 14 ± 1 | 9.7 ± 0.6 |
| WR8 | 3.1 ± 0.1 | 0.96 ± 0.04 | 23 ± 1 | 16 ± 1 |

**Table 5**

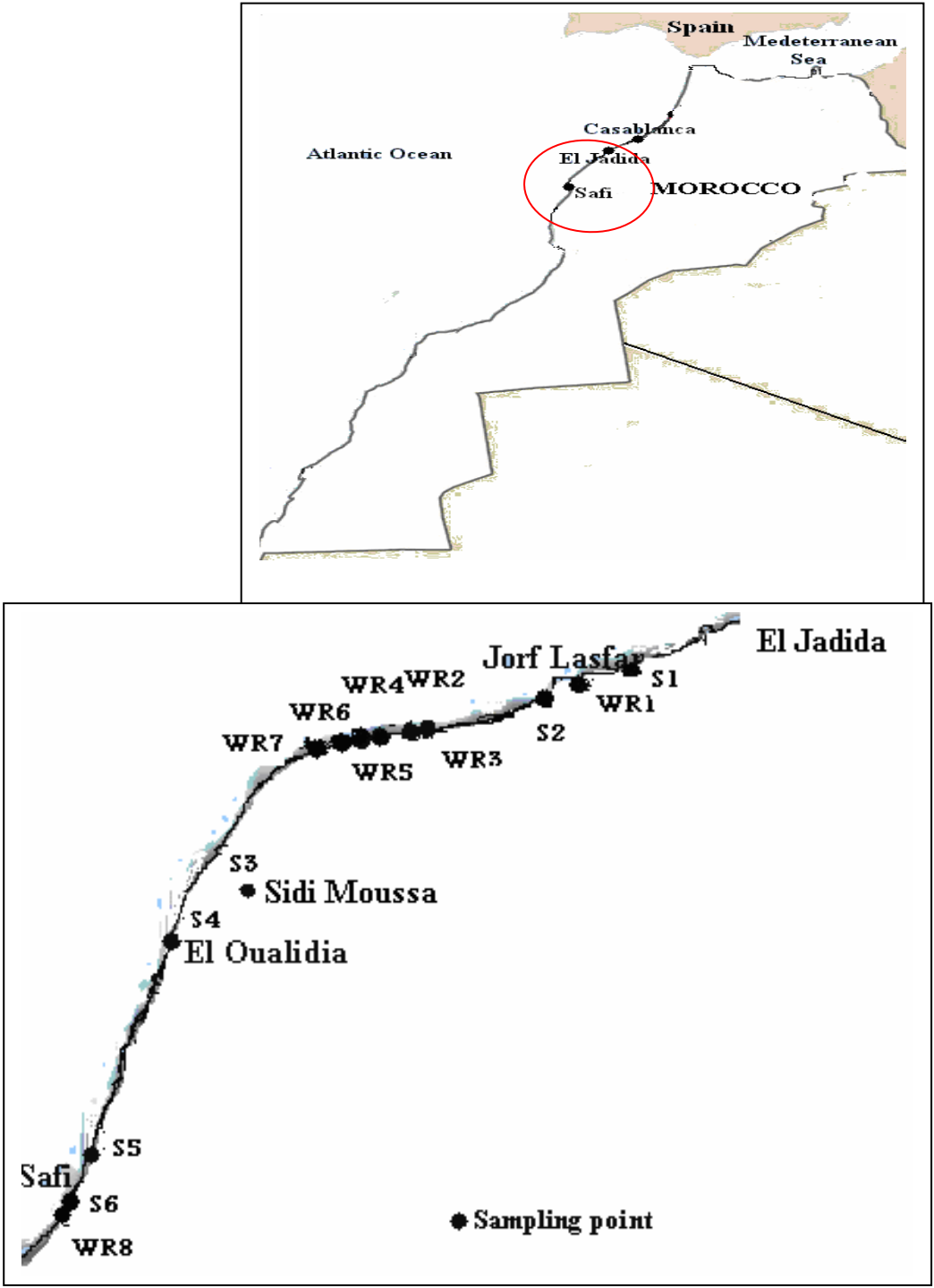

**Fig.1**

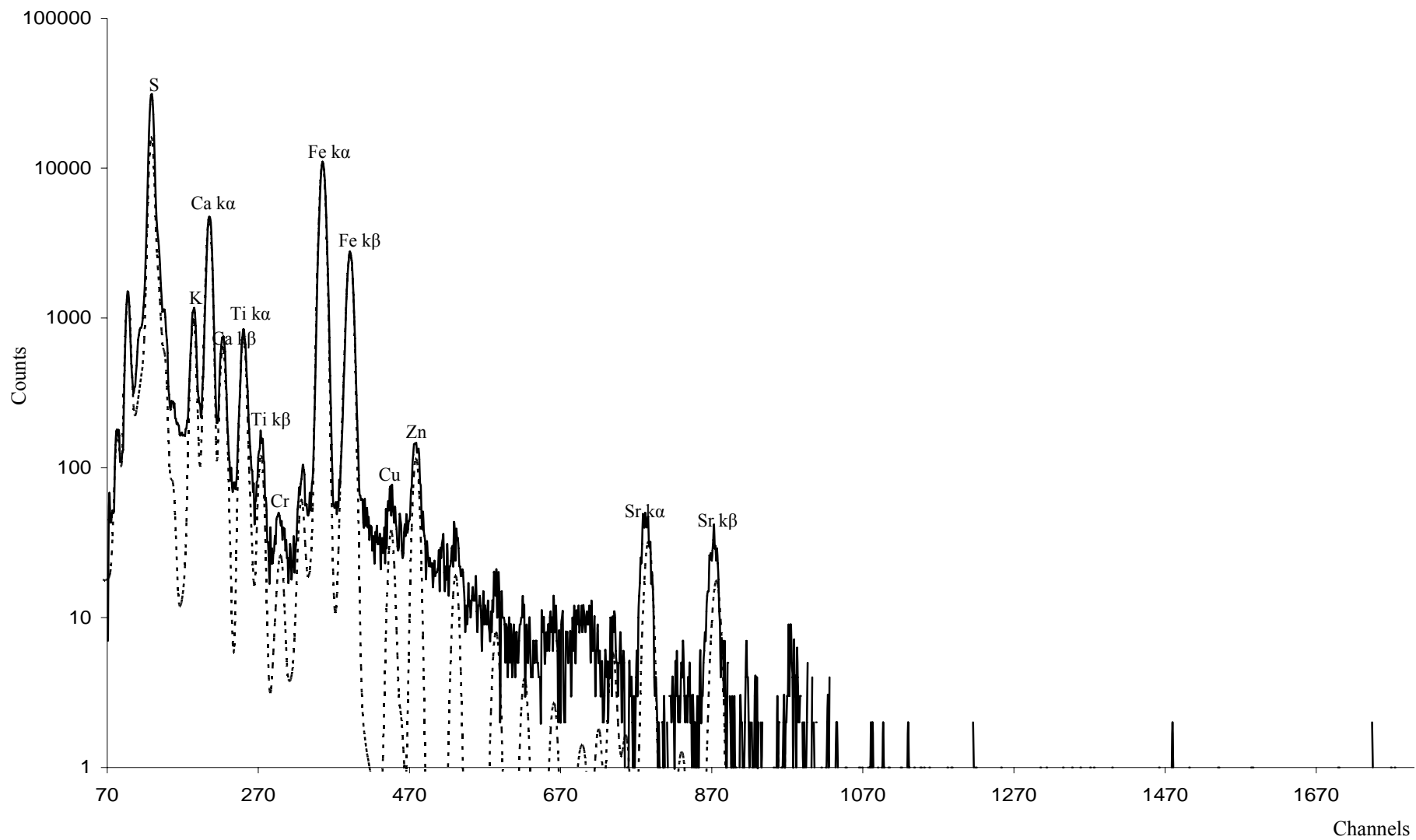

**Fig.2**

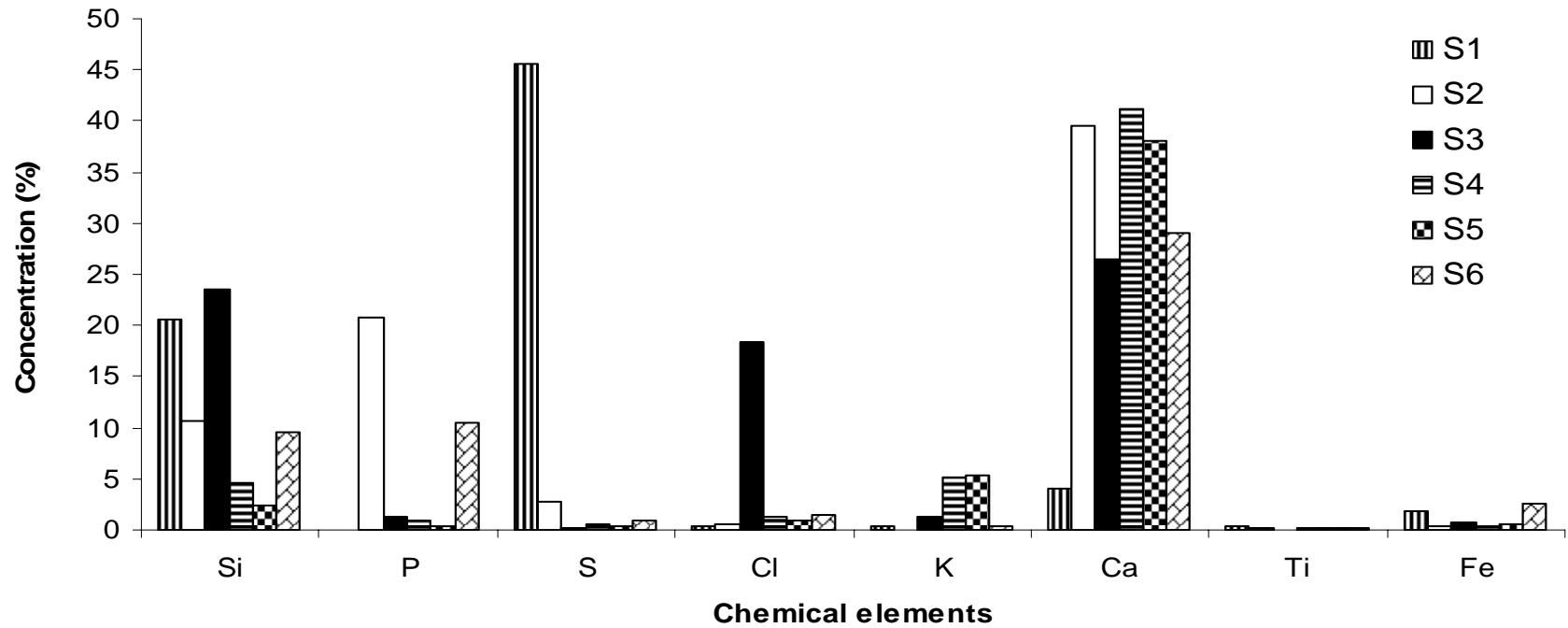

**Fig. 3(a)**

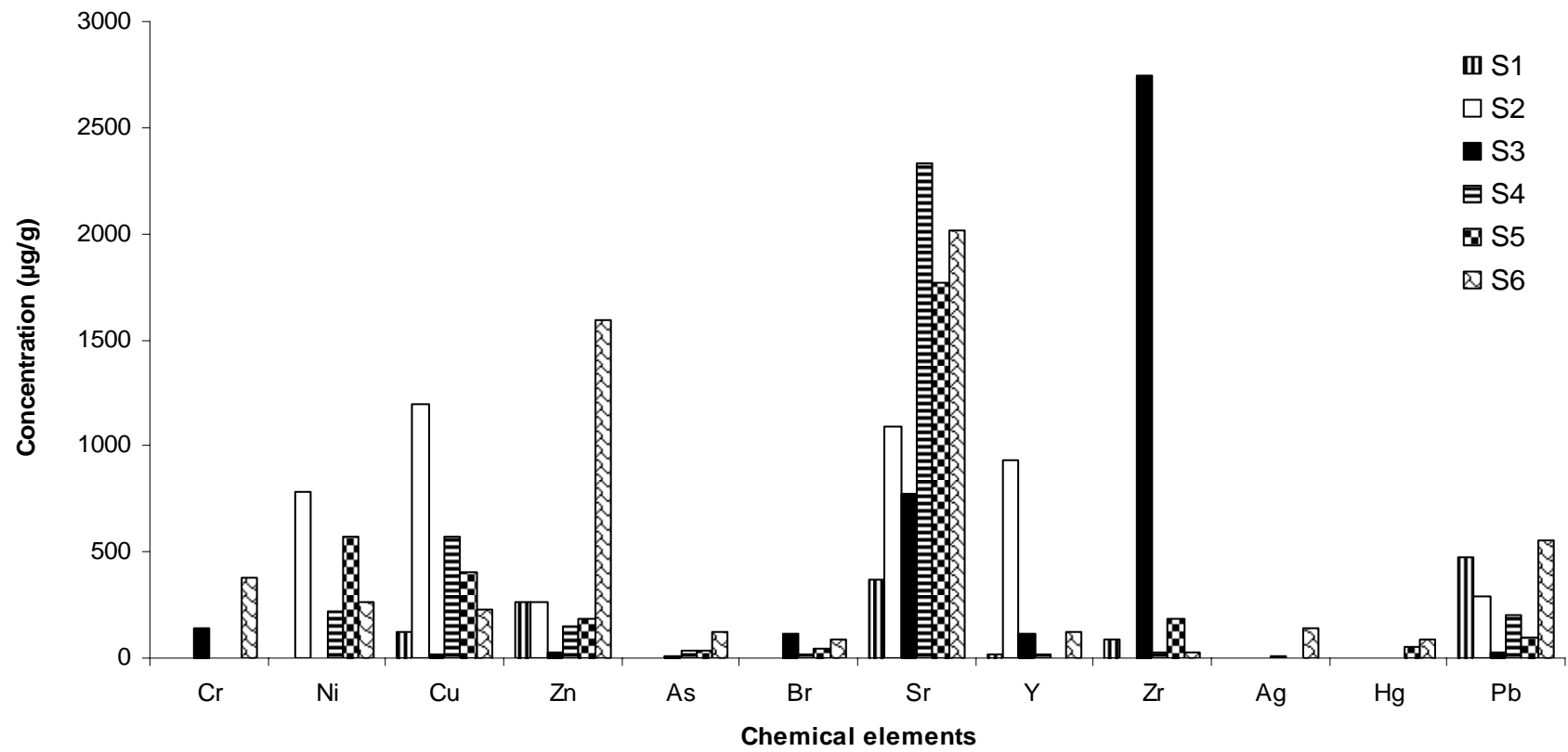

**Fig. 3(b)**

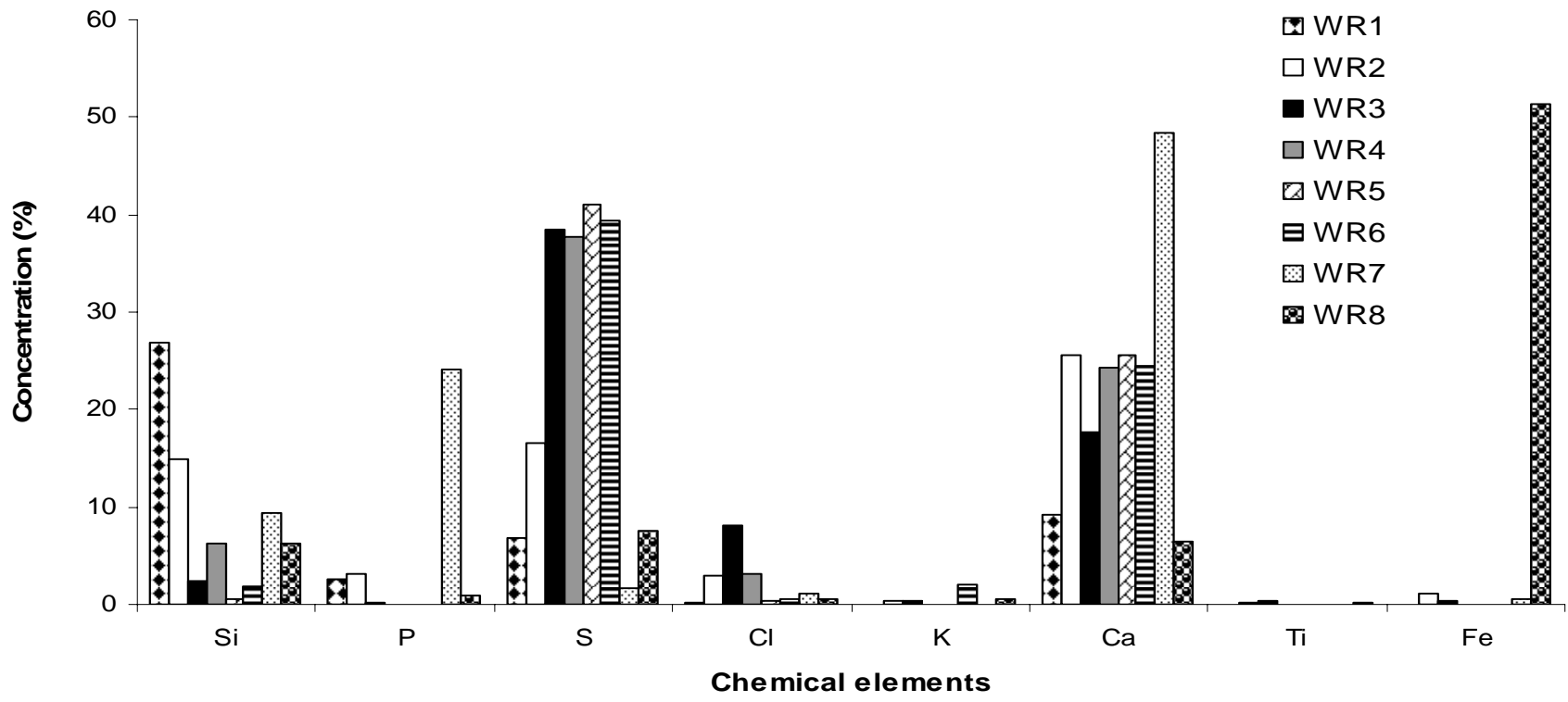

Fig. 4(a)

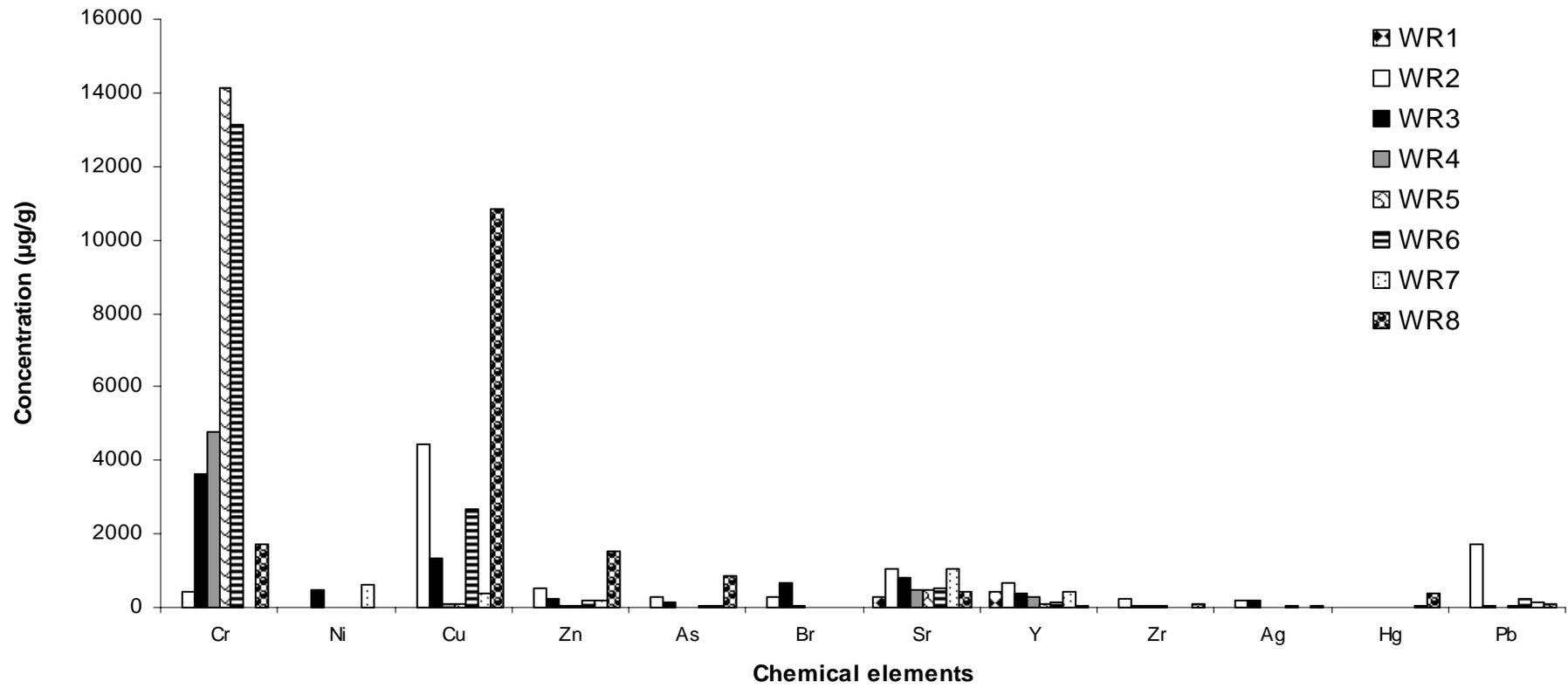

**Fig.4 (b)**